\documentclass[10pt,aps,prl,twocolumn,preprint,nofootinbib]{revtex4}

\usepackage{amsmath}    
\usepackage{graphicx}   
\usepackage{color}
\usepackage{verbatim}

\usepackage{cancel}

\begin{document}

\title{Tricritical Point in Quantum Phase Transitions of the Coleman-Weinberg Model at Higgs Mass}
\author{Miguel C N Fiolhais}
\email{miguel.fiolhais@cern.ch}   
\affiliation{LIP, Department of Physics, University of Coimbra, 3004-516 Coimbra, Portugal}
\author{Hagen Kleinert}
\email{h.k@fu-berlin.de}   
\affiliation{Freie Universit\"{a}t Berlin, Institut f\"{u}r Theoretische Physik Arnimallee 14, D-14195 Berlin, Germany}

\date{\today}
\begin{abstract} ~\\[-.5em]
The tricritical point, which separates first and second order phase transitions in three-dimensional superconductors, is 
studied in the four-dimensional Coleman-Weinberg model, and the similarities as well as the differences with respect to the three-dimensional result are exhibited.
The position of the tricritical point in the Coleman-Weinberg model is derived and found to be in agreement with the Thomas-Fermi approximation in the three-dimensional Ginzburg-Landau theory. From this we deduce a special role of the tricritical point for the Standard Model Higgs sector in the scope of the latest experimental results, 
which suggests the unexpected relevance of tricritical behavior in the electroweak interactions.
\\
\\
The following paper is published in Phys. Lett. A: http://www.sciencedirect.com/science/article/pii/S0375960113005768
\end{abstract}

\maketitle

\section{Introduction}
\label{sec:intro}
Ever since the formulation
 in 1964 of the electroweak spontaneous symmetry breaking mechanism \cite{ginzburg,anderson,englert,higgs,guralnik} to explain how elementary particles acquire mass, superconductivity and high-energy physics became intimately connected. For instance, the Ginzburg-Landau (GL) theory~\cite{ginzburg}, proposed in 1950 to provide a local macroscopic description of superconductivity, makes use of a quartic potential of the same type 
that reappeared in the Higgs model\footnote{With the short name ``Higgs'' we abbreviate collectively all authors that contributed from \cite{ginzburg,anderson,englert,higgs,guralnik}.}. 
In fact, the GL theory is the three-dimensional predecessor of 
what is now called 
(3+1)-dimensional scalar quantum electrodynamics,
that was studied in detail by 
Coleman and Weinberg \cite{coleman,weinberg}. Before the work of Ginzburg and Landau, the London theory
explained the existence of a finite penetration depth of magnetic fields into a superconductor, 
the Meissner-Ochsenfeld effect \cite{MO}.
GL 
extended this theory  by a local 
complex 
scalar order field $\phi(x)$, whose gradient terms in the energy density
produces a finite length scale of fluctuations of the order field,
the so-called coherence length $\xi$. 
In their theory, the Meissner-Ochsenfeld effect was explained by a
local 
mass term of the vector potential, 
whose size is proportional to $|\phi|^2$. 

The GL theory possesses two length scales, 
the London penetration depth $\lambda_L$, and the coherence length $\xi$.
The competition between the two is ruled by the GL parameter $\kappa\equiv \lambda_L/\sqrt{2}\xi$.
This serves to distinguish two types of superconductors, type-I with $\kappa>1/\sqrt{2}$ and 
 type-II with $\kappa<1/\sqrt{2}$.
The second type, possess bundles of vortices 
which confine 
magnetic flux in tubes of radius $\lambda_L$ 
 \cite{abrikosov}. 
In this way, the GL theory has become 
what may be called the {\it Standard Model\/} of superconductive 
phenomenology.

A similar Ginzburg-Landau-like scalar field theory with quartic interaction is 
successful in unifying the weak and the electromagnetic interactions, 
so that it has become the  {\it Standard Model\/} of particle physics, also
called the
{\emph{Higgs Model}}.  

An important new aspect that arises at the transition from the three-dimensional  
GL theory to the (3+1)-dimensional scalar electrodynamical Higgs model is that the field possess canonical 
commutation rules. These call for the existence of a
particle associated with each field.
After all, this 
is the logic which led to the discovery of pions as
the quantum of the forces of nuclear physics.
In particle physics,
it induced
an intensive search for a {\it Higgs particle\/} for many years. 
The recent discovery of a new signal in the 124-126 GeV mass region by the ATLAS and CMS collaborations at the Large Hadron Collider \cite{atlas,cms} 
is a hopeful candidate for such a particle.

\def\scy{superconductivity}
\def\sc{superconductive}

In this Letter we want to 
put this mass value into context with a   
known fact in \scy, that the \sc{} phase 
transition may occur in two different 
orders: a second order if the GL parameter $\kappa$ lies deep in
the type-II regime, and a first order in the type-I regime.
For a long time, this issue was a matter of
theoretical controversy after it had been 
argued by Halperin, Lubensky, and Ma (HLM) \cite{halperin}
that superconductors should really arise in a first-order transition.
The issue was finally settled by the calculation
of a tricritical point 
near the  dividing line between type-II to type-I superconductivity.
The approximate value of $\kappa$ where this happens was predicted 
to be $\kappa_{\rm tr}\approx 0.81/\sqrt{2}$ \cite{kleinert2,kleinertgauge,kleinert3},
a value later confirmed
by Monte Carlo simulations to lie 
at $\kappa=0.76/\sqrt{2} \pm 0.04$ \cite{hove}. 
The important point in the theory 
was that the mass term of the electromagnetic potential
was reliable only as long as it was big, which is the case in the type-I regime.
If it is small, the mass is destroyed by fluctuating vortex lines \cite{kleinert,kleinertbook}.
The  precise position of the 
tricritical point is unknown and should be determined by 
Monte Carlo simulations as described in \cite{MC}.

The calculation of HLM had an interesting parallel in (3+1)-dimensional scalar QED,
where Coleman and Weinberg\footnote{Their work was done almost simultaneously with \cite{halperin} on the same floor at Harvard University.} 
calculated that
a massless field would acquire a mass
from the fluctuations of the electromagnetic field. 
In the language of \sc, 
this implies that 
scalar QED has a first-order phase transition\footnote{On a hiking excursion into the mountains near Geneva with Sid Coleman, H.K. once asked him whether this was really what they proved, he said ``yes, but we foolishly did not put it that way''.}.
After the calculation of the tricritical point 
in \sc{} it was proposed 
that a similar tricritical point should 
come up in (3+1)-dimensional QED \cite{TCW}.
The Coleman-Weinberg result was derived 
without considering the fluctuating vortex sheets 
which are the (3+1)-dimensional analogs of 
the vortex lines in superconductors.
These should modify the CW-result in the small $e^2$ regime.
One should therefore expect a tricritical value of $\kappa$ also in (3+1)-dimensional
scalar QED, and the present Letter gives further support for this expectation with experimental consequences.
Moreover, the tricritical point is predicted and interpreted in the Standard Model as the absolute stability boundary of the Higgs potential, by analogy with superconductivity.
The latest theoretical predictions on the meta-stability and instability boundaries up to the Planck scale of the Standard Model Higgs potential are discussed in the context of the recent results on the observed signal at the LHC.

\section{Quartic Interaction and Tricritical Point}

The Ginzburg-Landau theory of superconductivity is characterized by the following energy density:
\begin{eqnarray}
 \mathcal{H} (\psi, {\nabla} \psi, \mathbf{A}, {\nabla} \mathbf{A}) & = &  \frac{1}{2} \left ( {\nabla} + i e \mathbf{A} \right ) \psi^* \left ({\nabla} - i e \mathbf{A} \right ) \psi \nonumber \\
&+& \frac{\tau}{2} |\psi|^2 + \frac{g}{4} |\psi|^4  \nonumber \\
&+& \frac{1}{2} \left ( {\nabla} \times \mathbf{A} \right )^2 \, ,
\label{hamiltonian}
\end{eqnarray}
with the order parameter $\psi(x) = \rho (x) \textrm{e}^{i\theta (x)}$, where $\rho(x)$ and $\theta (x)$ are real fields. 
The vector field is represented by $\mathbf{A}$, $e$ is the electric charge of the Cooper pairs\footnote{The Euler number is represented by $\textrm{e}$, and shall not be confused with the electric charge $e$.}, and the real constants $\tau$ and $g$ 
give the strength of the quadratic and quartic terms, respectively. If the mass parameter $\tau$ drops below zero, the ground state of the potential, 
$V(\psi) = \frac{1}{2} \tau |\psi|^2 + \frac{g}{4} |\psi|^4 $, is obtained for an infinite number of degenerate states satisfying:
\begin{equation}
\langle \psi \rangle^2 = \rho_0^2 = - \frac{\tau}{g} \, ,
\end{equation}
and corresponds to a second-order phase transition.
After the spontaneous symmetry breaking, \emph{i.e.} fixing the gauge to $\theta (x) = 0$, the Hamiltonian becomes,
\begin{eqnarray}
 \mathcal{H} (\psi, {\nabla} \psi, \mathbf{A}, {\nabla} \mathbf{A}) & = &  \frac{1}{2} \left ( {\nabla} \rho\right )^2 + V (\rho) + \frac{e^2\rho^2}{2} \mathbf{A}^2 \nonumber \\
&+& \frac{1}{2} \left ( {\nabla} \times \mathbf{A} \right )^2 \, .
\label{hamiltonian2}
\end{eqnarray}
The mass term of the vector field, $m_A = e \nu$, which appeared with the spontaneous symmetry breaking, 
and the scalar field mass term, can be associated with two characteristic lengths of a superconductor, 
the London penetration length, $\lambda_L = 1/m_A = 1/ e \rho_0  $, and the coherence length, $\xi = 1/\sqrt{-2\tau}$, respectively.

\begin{figure}[t]
\begin{center}
\includegraphics[height=4.cm]{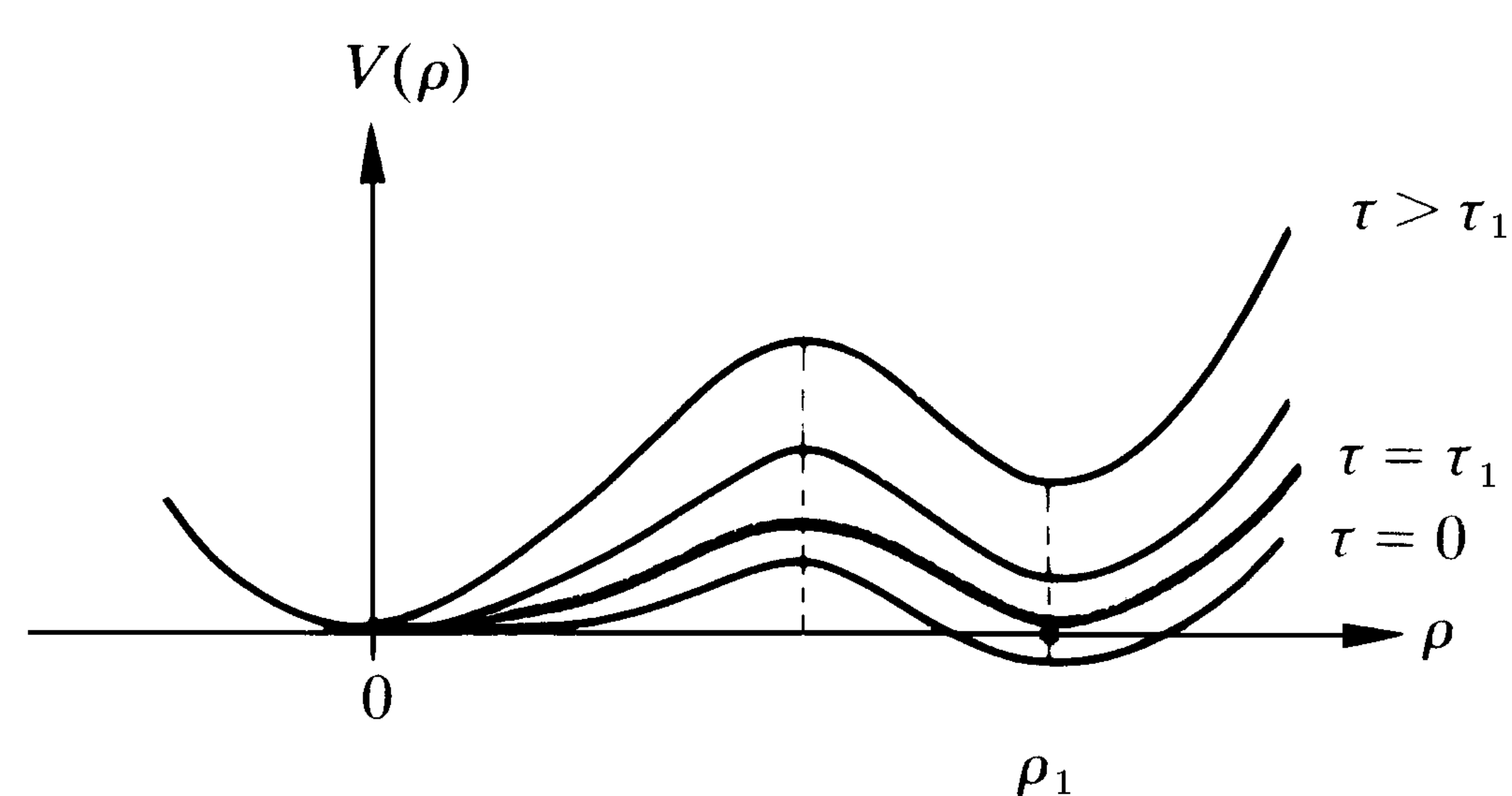}
\caption{Potential for the order parameter $ \rho $
with cubic term. A new minimum develops around $ \rho _1$
causing a first-order transition for $\tau =\tau _1$.}
\label{tricritical}
\end{center}
\end{figure}

The first-order phase transition can be achieved in the Ginzburg-Landau theory by considering quantum corrections, which in the
Thomas-Fermi approximation \cite{kleinertpath}, neglecting fluctuations in $\rho$, leads to an additional cubic term in the potential,
\begin{equation}
V(\rho) = \frac{1}{2} \tau \rho^2 + \frac{g}{4} \rho^4 - \frac{c}{3} \rho^3\, ,  \,\,\,\, c = \frac{e^3}{2\pi} \, .
\end{equation}
As shown in Fig. \ref{tricritical}, the cubic term generates a second minimum for $\tau < c^2/4g$, at the minimum,
\begin{equation}
\tilde{\rho}_0 = \frac{c}{2g} \left ( 1 + \sqrt{1-\frac{4\tau g}{c^2}} \right ) \, .
\end{equation}
At the specific point $\tau_1 = 2c^2/9g$, the minimum lies at the same level as the origin for $\rho_1 = 2c/3g$, where the phase transition becomes of first-order (tricritical point).
Therefore, in this point, the coherence length of the $\rho$-field fluctuations becomes,
\begin{equation}
\xi_1 = \frac{1}{\sqrt{\tau+3g\rho^2_1-2c\rho_1}} = \frac{3}{c} \sqrt{\frac{g}{2}} \, ,
\end{equation}
which is the same as the fluctuations around $\rho = 0$. Finally, the Ginzburg parameter at the tricritical point is,
\begin{equation}
\kappa = \frac{1}{2} \sqrt{\frac{g}{e^2}} \, .
\end{equation}

\section{Tricritical Point in the Coleman-Weinberg Model}

The intriguing question now is how this result changes in the four-dimensional version of the Ginzburg-Landau theory, the Coleman-Weinberg model.
The effective potential of the Coleman-Weinberg model at one-loop level is \cite{weinberg},
\begin{equation}
V(\phi_c) = \frac{1}{2} m^2 \phi_c^2 + \frac{\lambda}{4} \phi_c^4 + \frac{3e^4}{64\pi^2}\phi_c^4 \left ( \log{\frac{\phi_c^2}{M^2}} - \frac{25}{6} \right ) \, ,
\end{equation}
where the corrected scalar (spin-0) field is represented by $\phi_c (x)$, with a mass term $m^2$. Here $\lambda$ gives the strength of the quartic term,
and $M$ is the value of $\phi_c$ at which the renormalizations are done.
Note that we assumed $\lambda$ to be of the same order of $e^4$, and therefore, in the one-loop approximation,
the scalar loop diagrams were neglected, since they are of the same order of magnitude as the diagrams with two photon loops.
For convenience, a new variable $\mu$ can be defined as,
\begin{equation}
\frac{\lambda}{4} = \frac{3e^4}{64\pi^2} \left ( \log \frac{M^2}{\mu^2}  + \frac{11}{3} \right ),
\end{equation}
which turns the effective potential into,
\begin{equation}
V (\phi_c) = \frac{1}{2} m^2 \phi_c^2 + \frac{3e^4}{64\pi^2}\phi_c^4 \left ( \log \frac{\phi_c^2}{\mu^2}  - \frac{1}{2} \right ).
\end{equation}
As described in \cite{weinberg}, for a positive $m^2$, the effective potential has a maximum and minimum, for \mbox{$ m^2 < 3e^4\mu^2{\textrm e}^{-1}/16\pi^2 $}.
In particular, the minimum of the potential lies at the same level as the origin if \mbox{$m^2 = 3e^4\mu^2{\textrm e}^{-1/2}/32\pi^2$}, for $\langle \phi_c \rangle ^2 = \mu^2 {\textrm e}^{-1/2}$.
The mass of the scalar field in the tricritical point is, therefore,
\begin{eqnarray}
m^2 (\phi_c) & = & \left . \frac{\partial^2 V}{\partial \phi_c^2} \right |_{\phi_c  =  \langle \phi_c \rangle }   \nonumber \\ 
& = &  m^2  +  \frac{e^4 \langle \phi_c \rangle ^2}{16\pi^2} \left ( 6 + 9 \log{\frac{\langle \phi_c \rangle ^2}{\mu^2}} \right ) \nonumber \\
& = & \frac{3e^4\mu^2}{16\pi^2} {\textrm e}^{-1/2}  = \frac{\lambda}{\alpha} \langle \phi_c \rangle^2 ,
\end{eqnarray}
where $\alpha = \left ( \log \frac{M^2}{\mu^2}  + \frac{11}{3} \right )$ gives the size of the renormalization scale.
Consequently, at the tricritical point, the Ginzburg parameter becomes,
\begin{equation}
\kappa  =  \frac{1}{\sqrt{2}} \frac{1/e\langle \phi_c \rangle}{1 / m (\phi)}  =  \frac{1}{\sqrt{2\alpha}} \sqrt{\frac{\lambda}{e^2}}.
\end{equation}
The result has the same form as the previously obtained 3-dimensional result, 
and becomes the same with an appropriate choice of the renormalization scale.
Even though these results were computed using only the stability 
boundary of the corrected quartic potential, without making any use of the dual 
disorder field theory \cite{kleinertpath,kleinert2}, the position of the 
tricritical point does not change from 3 to 3+1 dimensions, thus justifying the applicability of 
the Thomas-Fermi approximation in the tricritical regime.

For the Standard Model Higgs potential, the relation between
the boundary of absolute stability and the tricritical point will be discussed in the next section.

\section{Higgs boson mass and vacuum stability}

On 4th July 2012, the CMS and ATLAS experiments announced the discovery of a new boson, compatible with the SM Higgs boson, with global statistical significances of 5.8 sigma (CMS) and 5.9 sigma (ATLAS). The observed signal currently lies at $125.3 \pm 0.4 \, (\textrm{stat.}) \pm 0.5 \, (\textrm{sys.})$~GeV (CMS) and 
$126\pm0.4 \, (\textrm{stat.}) \pm 0.4 \, (\textrm{sys.})$~GeV (ATLAS), and no significant deviations from the predicted SM Higgs boson properties were observed to the present date.

Assuming the Standard Model to be valid up to the Planck scale, the Higgs potential develops a new local minimum for a positive value of the running quartic coupling with the renormalization scale. However, for a negative quartic coupling, the potential becomes unbounded from below and, therefore, unstable. Thus
 the absolute stability of the electroweak vacuum has its boundary where the quartic coupling flips sign. This feature of the SM Higgs potential corresponds precisely to the 
previously discussed tricritical point in a quartic interaction, which separates first and second order phase transitions in superconductors. The phenomenology associated with the two physical situations is, of course, quite different. While in superconductivity, 
the spontaneous symmetry breaking appears as a result of the radiative corrections overcoming the effect of a positive mass term in the Coleman-Weinberg model, 
the SM Higgs potential is characterized by the existence of two non-zero vacua. Nevertheless, it is clear that the vanishing of the quartic coupling and the degeneracy of the vacuum states correspond to a tricritical behavior in the two scenarios.

The determination of the SM vacuum stability has been studied in detail in the past two decades \cite{smvacuum}.
The latest and most precise next-to-next-to-leading-order (NNLO) prediction of the absolute stability boundary 
was established by Degrassi \emph{et al.} \cite{degrassi}, using two-loop renormalization-group equations, one-loop threshold corrections at the electroweak scale 
(possibly improved with two-loop terms in the case of pure QCD corrections), and one-loop
improved effective potential. Assuming a top quark mass of $m_t = 173.1 \pm 0.7$~GeV \cite{top}, and the strong coupling constant at $\alpha_s (M_Z) = 0.1184 \pm 0.0007$ \cite{strong}, 
the absolute stability boundary up to the Planck scale was predicted for a Higgs boson mass,
\begin{eqnarray}
m_H[\textrm{GeV}]  & = & 129.4 \nonumber \\ 
                                    & \pm & 1.4 \left ( \frac{m_t[\textrm{GeV}]-173.1}{0.7} \right ) \nonumber \\
                                    & \pm & 0.5 \left ( \frac{\alpha_s (M_Z)-0.1184}{0.0007} \right ) \nonumber \\ 
                                    & \pm & 1.0~{\textrm{(theoretical)}} \, .
\end{eqnarray}
By combining in quadrature the theoretical and experimental uncertainties, the result becomes $m_{H} = 129.4 \pm 1.8~\textrm{GeV}$, 
distanced by roughly 2 sigma from the LHC results. Therefore, one cannot state that the Higgs boson lies precisely on the 
tricritical point of the electroweak interactions, nor exclude that possibility. 
The allowed regions, up to 3 sigma, on the top quark and Higgs boson masses measurements, seem to indicate a significant preference for meta-stability of the SM potential
when compared with the latest experimental results from the Tevatron and LHC.
This tells us there is a non-zero probability of quantum tunneling into the global minimum, lying deeper than the electroweak vacuum. 
As the new vacuum appears at a very high-energy scale, the probability of tunneling is very small, with a mean lifetime larger than the age of the Universe.
Nonetheless, the absolute stability boundary strongly depends on the Higgs boson and top quark masses: slight variations may have dramatic implications.
The possible improvement of the precision of these observables at the LHC and in future linear colliders, and further progress on the 
theoretical understanding of the vacuum stability, may provide further insights into the nature of the Higgs boson mass.

All of this assumes, of course, that the Standard Model is valid all the way up to the Planck scale. 
So far, the observed data at the LHC has been found to be in agreement with the Standard Model predictions.
However, there is no obstacle that would prevent
the existence of new physics contributing
at higher energy scales, beyond the current reach of the LHC, and this
could well
 have an impact on the stability of the Higgs potential. These new physics effects, above the electroweak symmetry breaking scale, can be parameterized in a 
model-independent way by an effective 
Lagrangian~\cite{Buchmuller:1985jz},
\begin{equation}
\mathcal{L} = \mathcal{L}_\text{SM} + \sum \frac{C_x}{\Lambda^2} O_x + \dots \,,
\label{ec:effL}
\end{equation}
where $O_x$ are dimension-six operators invariant under the SM gauge
symmetry, $C_x$ are the dimensionless operator coefficients,
 and $\Lambda$ is the new physics scale.
The effect of such operators has been studied in the past~\cite{newphysics} and shown to have a significant influence on the stability and triviality of the Higgs potential. For instance, new physics contributions at an energy scale of a few TeV could be enough to ensure the stability of the electroweak vacuum. Perhaps the quest for anomalous contributions to the top quark and Higgs boson SM couplings at the LHC may bring us interesting surprises in the years ahead~\cite{anomalous}.

\section{Summary}

In this Letter, we argue that the tricritical point, obtained for the three-dimensional Ginzburg-Landau theory with the help of a duality transformation to a disorder version, 
is not expected to significantly change when analyzed in the context of the (3+1)-dimensional Coleman-Weinberg model.
This leads us to conclude that the absolute stability boundary of the Higgs potential is a tricritical point of the electroweak interaction, by analogy with superconductivity.
The recently obtained result on the NNLO prediction of the absolute stability boundary, up to the Planck scale, at a Higgs boson mass of $m_{H}\approx129.4\pm1.8~\textrm{GeV}$, 
compatible with the observed signals at LHC in the 124-126 GeV mass region, suggests that the electroweak interactions make use of the tricritical behavior as its natural working point.
To validate this statement, we must wait for a greater precision of the experimental measurements and theoretical predictions.
Finally, and more strategically, this interpretation may 
enhance the bridge between the physics of 
elementary particles and 
superconductivity, that has led to many important insights
since Nambu's pioneering work on the chiral phase transition.



\end{document}